\documentclass[manuscript]{aastex}
\usepackage{rotating,graphicx}

\shorttitle{Solar Cycle Variations of Solar Wind Turbulence
Anisotropy}

\shortauthors{Zhou \& He}

\begin{document}

\title{The Solar Cycle Variations of the Anisotropy of Taylor
Scale and Correlation Scale in the Solar Wind Turbulence \\}

\author{G. Zhou\altaffilmark{1,2} and H.-Q. He\altaffilmark{1,3,4}}

\altaffiltext{1}{Key Laboratory of Earth and Planetary Physics,
Institute of Geology and Geophysics, Chinese Academy of Sciences,
Beijing 100029, China; hqhe@mail.iggcas.ac.cn}

\altaffiltext{2}{College of Earth and Planetary Sciences, University
of Chinese Academy of Sciences, Beijing 100049, China}

\altaffiltext{3}{Innovation Academy for Earth Science, Chinese
Academy of Sciences, Beijing 100029, China}

\altaffiltext{4}{Beijing National Observatory of Space Environment,
Institute of Geology and Geophysics, Chinese Academy of Sciences,
Beijing 100029, China}

\begin{abstract}
The field-aligned anisotropy of the solar wind turbulence, which is
quantified by the ratio of the parallel to the perpendicular
correlation (and Taylor) length scales, is determined by
simultaneous two-point correlation measurements during the time
period 2001-2017. Our results show that the correlation scale along
the magnetic field is the largest, and the correlation scale in the
field-perpendicular directions is the smallest, at both solar
maximum and solar minimum. However, the Taylor scale reveals
inconsistent results for different stages of the solar cycles.
During the years 2001-2004, the Taylor scales are slightly larger in
the field-parallel directions, while during the years 2004-2017, the
Taylor scales are larger in the field-perpendicular directions. The
correlation coefficient between the sunspot number and the
anisotropy ratio is employed to describe the effects of solar
activity on the anisotropy of solar wind turbulence. The results
show that the correlation coefficient regarding the Taylor scale
anisotropy (0.65) is larger than that regarding the correlation
scale anisotropy (0.43), which indicates that the Taylor scale
anisotropy is more sensitive to the solar activity. The Taylor scale
and the correlation scale are used to calculate the effective
magnetic Reynolds number, which is found to be systematically larger
in the field-parallel directions than in the field-perpendicular
directions. The correlation coefficient between the sunspot number
and the magnetic Reynolds number anisotropy ratio is -0.75. Our
results will be meaningful for understanding the solar wind
turbulence anisotropy and its long-term variability in the context
of solar activity.
\end{abstract}


\clearpage

\section{Introduction}
Plasma turbulence is a common phenomenon occurring in nature and the
turbulence in the heliosphere plays a significant role in several
aspects of space plasma behaviors, such as high-energy particle
acceleration, solar wind generation, plasma heating, galactic cosmic
ray modulation, and solar energetic particle propagation
\citep{1965PhFl....8.1385K,1971ApJ...168..509B,1982JGR....87.6011M,1982JGR....8710347M,1995SSRv...73....1T,2016JPlPh..Chen,He2019}.
Recently, the anisotropy has become one of the important aspects of
the investigations of the solar wind, especially of the solar wind
turbulence \citep[e.g.,][]{2012SSRv_Horbury}. The turbulence
anisotropy may affect the acceleration and transport of energetic
particles, the heating of plasmas, and the propagation of cosmic
rays in the heliosphere
\citep{1968ApJ...152..799J,1968ApJ...152..997J,1970ApJ...160..745J,2003PPCF...Velli,2005PPCF_Duffy,He2015}.

One task in the field of solar wind turbulence anisotropy is to find
a theoretical model for describing the turbulence anisotropy. In the
solar wind observations, there are several types of fluctuation
anisotropy, e.g., variance anisotropy, energy transfer rate
anisotropy, power anisotropy, and correlation anisotropy (also known
as spectral anisotropy or wavevector anisotropy). Of most interest
is the correlation anisotropy
\citep{2005PPCF_Duffy,2007ApJ...654L.103O}. At present, there exist
three models for the correlation anisotropy, i.e., the ``slab"
model, the two-dimensional (2D) model, and the slab+2D composite
model. According to the slab model, the correlation function decays
in the directions parallel to the mean magnetic field, but without
field-perpendicular variations. This means that the correlation
function has the shortest scales in the directions parallel to the
mean magnetic field and the longest scales in the
field-perpendicular directions. In contrast, the 2D model considers
that the correlation function decays only in the directions
perpendicular to the mean magnetic field. The slab+2D two-component
model was proposed by \citet{1990JGR....Matthaeus}. They found that
the turbulence is not consistent with either the slab model or the
2D model, but shows a ``Maltese cross'' shape. Therefore, they used
a superposition of the slab model and the 2D model to interpret this
phenomenon. \citet{1996JGR...Bieber} took the ratio of the
perpendicular to quasi-parallel power spectra to quantify the
anisotropy of the turbulence, and showed that near $80\%$ of the
energy is in the 2D component and $20\%$ in the slab component.
Further, \citet{2005ApJ...Dasso} found that in the fast solar wind
and at the larger scale of the inertial range, the correlation
scales are longer in the field-perpendicular directions than in the
field-parallel direction (slab model), whereas for the slow solar
wind situation, the 2D component is predominant. Later, a similar
result was obtained by \citet{2011JGRA..116.8102W}. Despite the
two-component model is a rather idealized model with drastic
approximation, it can present the dominant properties of the solar
wind turbulence and provide a useful parameterization for the
anisotropy studies
\citep{1990JGR....Matthaeus,1996JGR...Bieber,2005NPGeo..Oughton,2005ApJ...Dasso,2007ApJ...654L.103O,2009JGRA..114.7213W,2011JGRA..116.8102W,2012SSRv_Horbury}.

In most studies, a consensus conclusion can be found: the
correlation length scales of the turbulent eddies in the directions
parallel to the local mean magnetic field ${{\textbf{B}}_0}$ are
much larger than those in the directions perpendicular to
${{\textbf{B}}_0}$. This indicates that the anisotropy of the
correlation length scales is dominated by the 2D model. However,
this is not the case for the dissipation scales. As the energy
cascade proceeds to smaller scales, and reaches the plasma
micro-scales, such as ion cyclotron (inertial) scales, electron
cyclotron (inertial) scales, and possibly even smaller scales, the
dissipation and heating processes are usually thought to take place.
At dissipation scales, the observations of correlation anisotropy
appear to be a little confusing \citep{2016JPlPh..Chen}. On one
hand, \citet{2009JGRA..114.7213W} found that the Taylor scale, which
is related to the dissipation scale
\citep{Tennekes1972A,2005JGRA..110.1205W}, is independent of the
directions relative to the mean magnetic field. On the other hand,
some studies showed that the slab component dominates the anisotropy
near the dissipation scales
\citep{2008JGRA..Hamilton,2011_JGR_Tessein,2012ApJ...Smith}, while
others argued that the 2D component remains dominant
\citep{2010PhysRevLett_Sahraoui,2010PhysRevLett_Chen,2011GeoRL..Narita,2014npg_Comisel,2014ApJ...Perschke}.
\citet{2009ApJ...Podesta} also pointed out that the anisotropy near
dissipation scales exhibits an even more complex behavior. The
reason leading to this phenomenon might be that different studies
may use different techniques, and the distinct scale ranges can
influence the results as well. Another possible reason, which will
be tested in this work, is that the Taylor scale is apt to be
affected by the solar wind conditions in the context of long-term
variations.

Most of the previous work has been done with single-spacecraft
measurements and the frozen-in flow assumption was usually used when
the mean velocity was supersonic and super-Alfv{\'e}nic
\citep{1990JGR....Matthaeus,1996JGR...Bieber,2005NPGeo..Oughton,2007ApJ...654L.103O}.
This situation has to some degree been improved in recent years due
to the increasing exploration missions in the near-Earth solar wind.
Multi-spacecraft analyses that can directly characterize the
turbulence anisotropy without invoking the frozen-in approximation
have achieved significant progress by employing modern techniques
such as simultaneous two-point correlation functions
\citep{2005PhRvL..95w1101M}. In this work, we examine the
correlation anisotropy in the solar wind fluctuations during January
2001 to December 2017, which covers an entire solar cycle. The
effects of solar activity on the turbulence anisotropy are
investigated. As usual, the anisotropy is quantified by the ratio of
the field-perpendicular to the field-parallel correlation (Taylor)
length scales. This paper is structured as follows. In Section
\ref{sec:method}, we provide a detailed description of the method
and procedure of the two-point measurements. In Section
\ref{sec:results}, we calculate the correlation (Taylor) length
scale in each angular bin during different time ranges, and discuss
how the solar activity influence the anisotropy of the space plasma
turbulence. A summary of our results will be provided in Section
\ref{sec:summary}.

\section{Methods and Procedure} \label{sec:method}
In the simple scenario of homogeneous turbulence, the means,
variances, and correlation values of the fluctuations should be
independent of the choice of the coordinate system origin
\citep{1953tht..book.....B,Tennekes1972A,1979sswp.book..249B,2000ifd..book.....B}.
For a magnetic field ${\textbf{B}}({\textbf{x}},t) =
{{\textbf{B}}_0} + {\bf{b}}$, the mean is $\left\langle {\textbf{B}}
\right\rangle  = {{\textbf{B}}_0}$, the fluctuation is ${\textbf{b}}
= {\textbf{B}} - {{\textbf{B}}_0}$, and the variance is ${\sigma^2}
= \left\langle {{{\left| {\textbf{b}} \right|}^2}} \right\rangle$.
The two-point correlation coefficient is
\begin{equation}
R({\bf{r}}) = \frac{1}{\sigma^2}\left\langle {{\bf{b}}({\bf{x}})
\cdot {\bf{b}}({\bf{x}} + {\bf{r}})} \right\rangle. \label{cor}
\end{equation}
Here $\textbf{r}$ is the separation between the two points
$\textbf{x}$ and ${\textbf{x}} + {\textbf{r}}$. For homogeneity, $R$
and $\textbf{B}_0$ are independent of $\textbf{x}$. The
$\left\langle {...} \right\rangle$ denotes an ensemble average,
which is determined by a suitably chosen time-averaging procedure.
The direction-averaged correlation scale is defined as
\citep{2005PhRvL..95w1101M}
\begin{equation}
L = \int_0^\infty{R(r)}dr. \label{L}
\end{equation}
Therefore, an ansatz function form of ${R(r)}$ can be derived from
Equation (\ref{L}) as $R(r)\sim{e^{-r/L}}$, thus $R(r)=1$ for $r=0$
and $R \to 0$ for $r \to \infty$. Similarly, the Taylor scale
$\lambda$ can be determined from ${R(r)}$ as well. The Taylor scale
is the length scale associated with the second order expansion of
the two-point correlation function ${R(r)}$ evaluated at zero
separation. That is to say, the Taylor scale is the radius of
curvature of the correlation function at the origin, and can be
obtained from the expression $R(r)\sim 1 - {(r/\sqrt{2}\lambda)^2}$
(for more details, see \citet{2005PhRvL..95w1101M} and
\citet{2007JGRA..11210201W}). With these definitions, the effective
magnetic Reynolds number can be obtained from the following
expression
\begin{equation}
R_m^{eff} = {\left(\frac{L}{\lambda}\right)^2}. \label{R_m}
\end{equation}

The magnetic field data used in this investigation were measured by
the triaxial fluxgate magnetometers on board spacecraft Advanced
Composition Explorer (ACE), Wind, and Cluster during the time period
from January 2001 to December 2017. In our previous work, we have
presented the procedures for determining ${R(r)}$, $L$, $\lambda$,
and $R_m^{eff}$ from multi-spacecraft data, and have shown some
novel results of these scales during an entire solar cycle
\citep{2020ApJ...899L..32Z}. In this work, we use sufficient samples
to resolve the correlations into angular bins that deviate from the
direction parallel to the mean magnetic field, and further
investigate the relationship between the correlation anisotropy and
the solar activity.

In the analyses, we interpolate the spacecraft data (ACE-Wind and
Cluster) to 1 minute resolution to obtain the field vectors at
different positions. Note that the Cluster's orbit was in proximity
to the magnetosphere. Therefore, the foreshock waves were sometimes
present in the solar wind measurements. To reduce the influences of
the foreshock waves, in our investigations the magnetic field
measurements from the Cluster spacecraft in the solar wind are
averaged to 1 minute resolution, which is much longer than the
longest period for ion foreshock waves (approximately 30 seconds).
Similar procedure for processing Cluster data was adopted by
\citet{2007JGRA..11210201W,2011JGRA..116.8102W}. The individual
correlation estimates from the ACE-Wind data are computed by
averaging over 24-hr contiguous intervals of measurements. However,
the correlation estimates from the Cluster data are calculated by
averaging over 2-hr intervals of measurements, since the Cluster
spacecraft do not remain in the solar wind for long periods. The
entire data set during 2001-2017 is divided into 15 3-yr time
periods (2001-2003, 2002-2004, $\ldots$, 2014-2016, 2015-2017). In
each 3-yr time period, the data intervals (approximately 500 24-hr
intervals for ACE-Wind data and 1000 2-hr intervals for Cluster
data) are randomly selected for calculating the correlation scales
and the Taylor scales. For a detailed description of the technical
procedures, we refer the reader to our previous work of
\citet{2020ApJ...899L..32Z}. In each data interval, we calculate the
mean magnetic field by averaging the magnetic field vector over the
entirety of the data interval, and further compute the magnetic
field vector's time-averaged two-point correlation coefficients.
This value is assigned to the time-averaged separation distance
between the corresponding two spacecraft in the corresponding
interval. We then determine the average value of these separation
values along the field-parallel directions (X axis) and the
field-perpendicular directions (Y axis) with respect to the mean
magnetic field, and assign it into the angular bin. Every angular
bin is ${10^\circ }$ wide except the bin near the field-parallel
direction, which is set to be ${30^\circ}$ wide (due to the
consideration of the fewer intervals close to the parallel direction
of the mean magnetic field). This angular division scheme is adopted
from \citet{2009JGRA..114.7213W,2011JGRA..116.8102W}. We note that
using width of ${20^\circ}$ or ${10^\circ}$ for the angular bin near
the field-parallel direction will not qualitatively affect the
obtained values for the correlation scales. Nevertheless, using the
width of ${20^\circ}$ or ${10^\circ}$ for the angular bin near the
field-parallel direction will lead to some negative (invalid) values
of the obtained Taylor scales therein, due to the limited valid data
in the angular bin. Therefore, we decide to use the width of
${30^\circ}$ for the first bin (near the field-parallel direction),
as in \citet{2009JGRA..114.7213W,2011JGRA..116.8102W}.

\begin{figure}
 \epsscale{0.5}
 \plotone{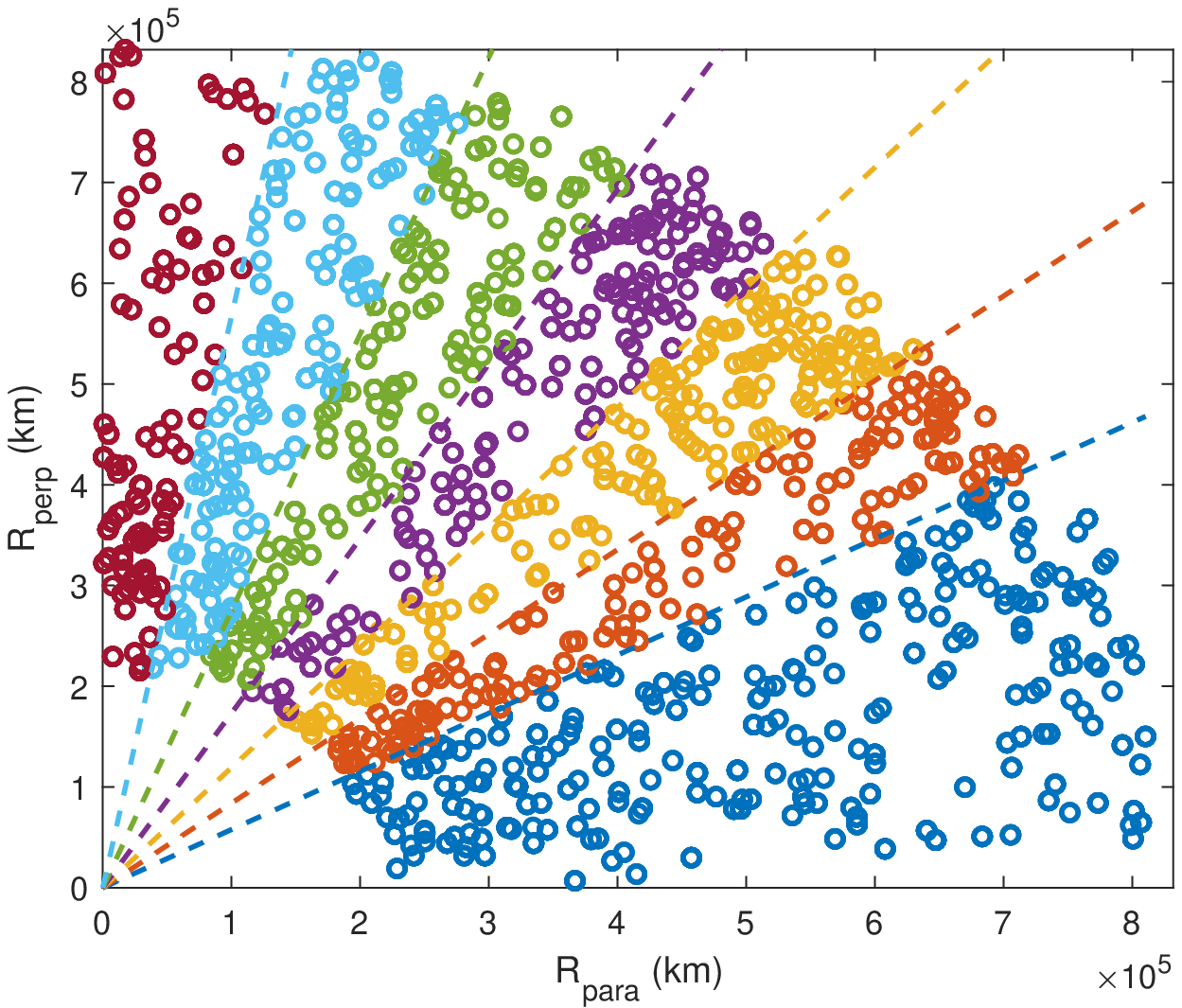}
 \plotone{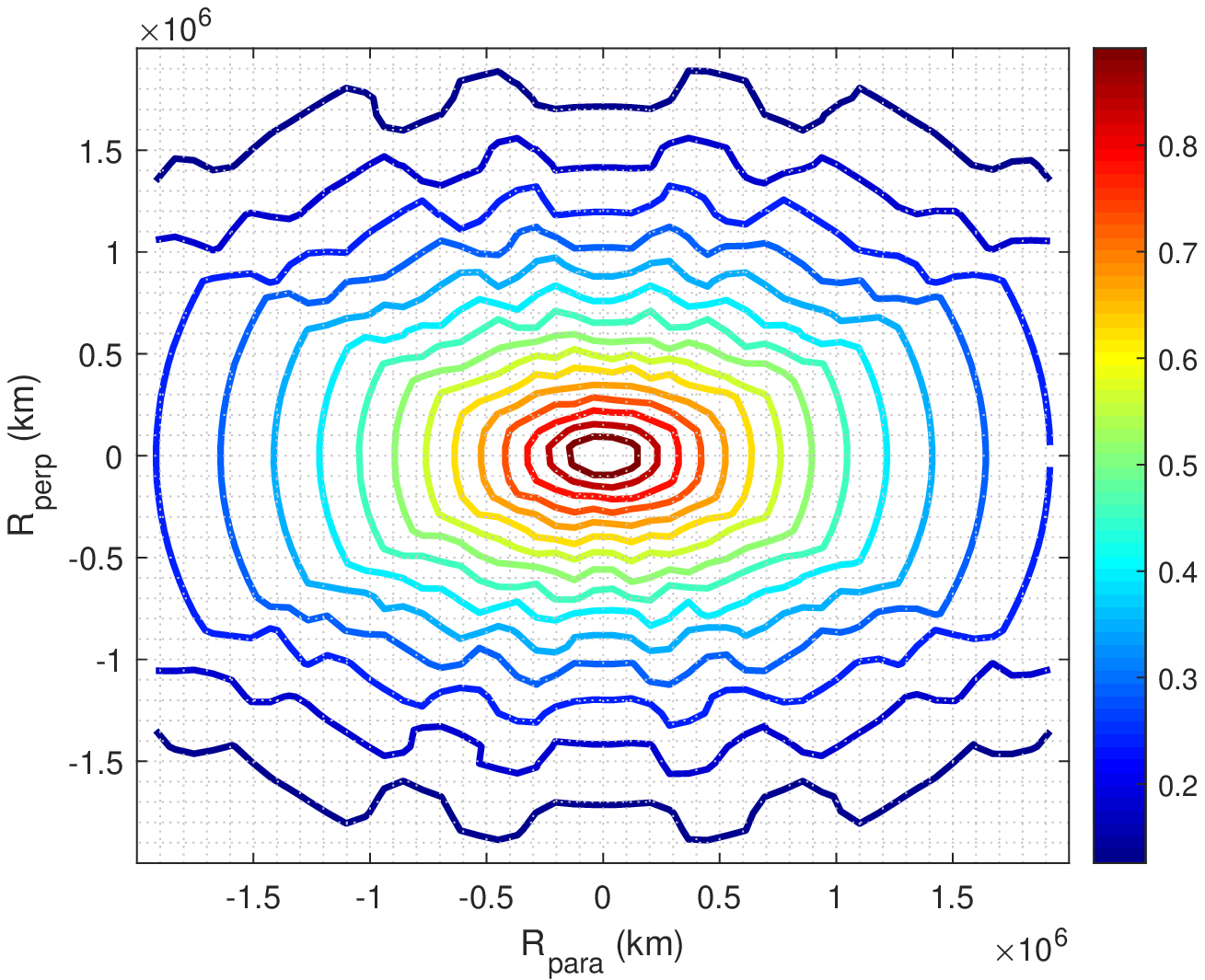}
 \caption{Top: distribution of the spacecraft separation distances in the field-parallel
 and field-perpendicular directions during the years 2005-2007. The dashed lines denote the
 boundaries of the angular bins, and the circles denote the spacecraft separations
 within the corresponding angular bins. Bottom: contour plot for the correlation
 function during the time period 2005-2007. The color scale bar denotes the values of the
 corresponding correlation coefficients. \label{samples}}
\end{figure}

As an example, the top panel of Figure \ref{samples} shows the
distribution of the data of the separation distances in the
directions parallel to and perpendicular to the mean magnetic field
during the divided time period 2005-2007, and the bottom panel of
Figure \ref{samples} displays the corresponding plot of the
correlation function contour. As we can see, the values of the
correlation coefficients vary more sharply in the
field-perpendicular directions than in the field-parallel
directions, which indicates that the correlation length scales are
the largest along the mean magnetic field and the smallest in the
field-perpendicular directions (approximately 2D-dominant). The
ratio of the field-parallel to the field-perpendicular correlation
length scales is 1.63, which is similar to the result of $1.79 \pm
0.36$ proposed by \citet{2007ApJ...654L.103O}.

We obtain all the values of the correlation scale and the Taylor
scale in different angular bins during each divided time period, and
calculate the ratio of the field-parallel to the field-perpendicular
values. In this work, we use the sunspot number as the indicator of
the solar activity and solar cycle
\citep{1979cmft.book.....P,2010LRSP....7....1H}. Then we can
investigate the solar cycle variability of the anisotropy of the
solar wind turbulence.

\section{Results and Discussion} \label{sec:results}

\subsection{Results}
\begin{figure}
 \epsscale{0.5}
 \plotone{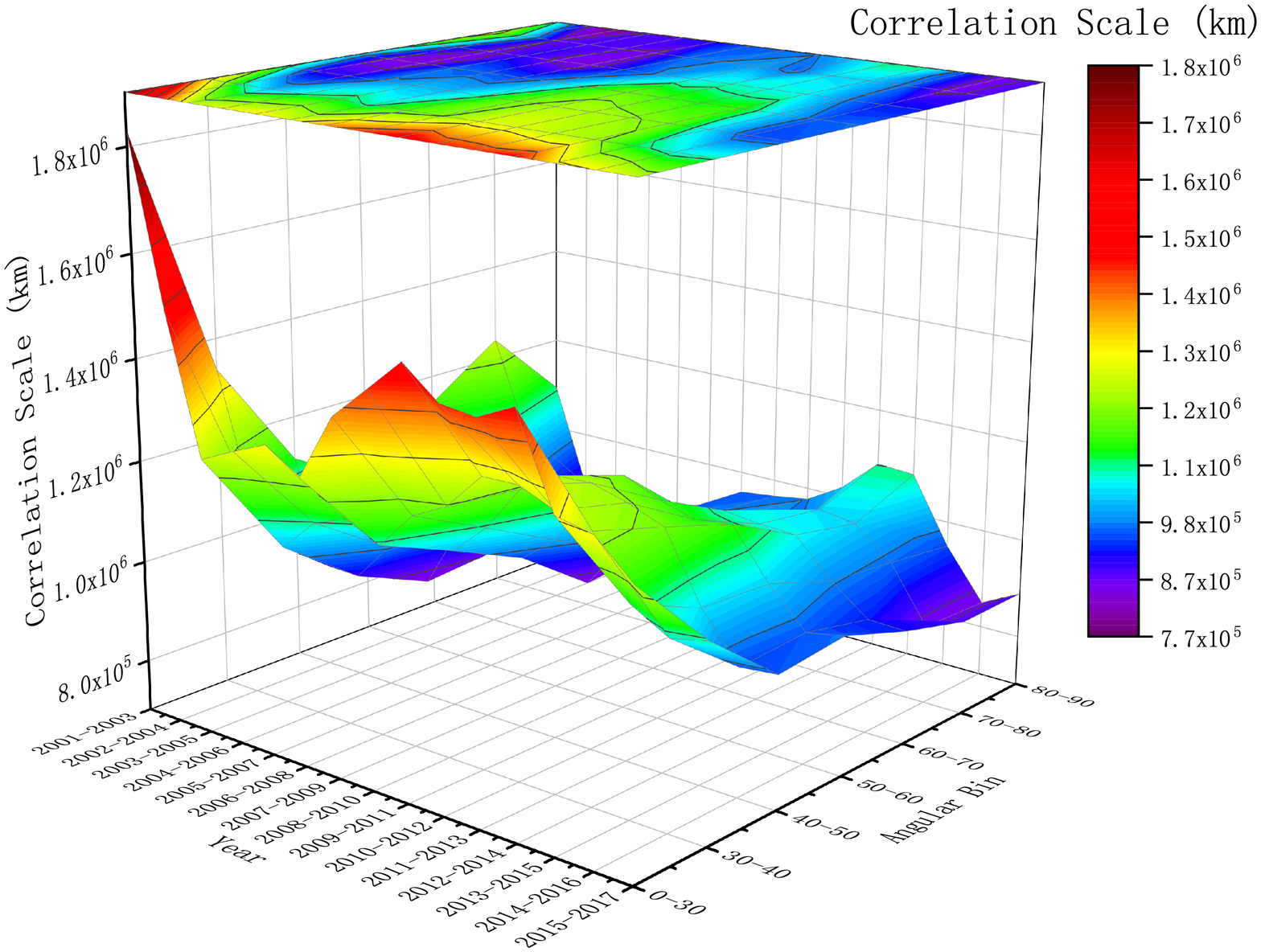}
 \plotone{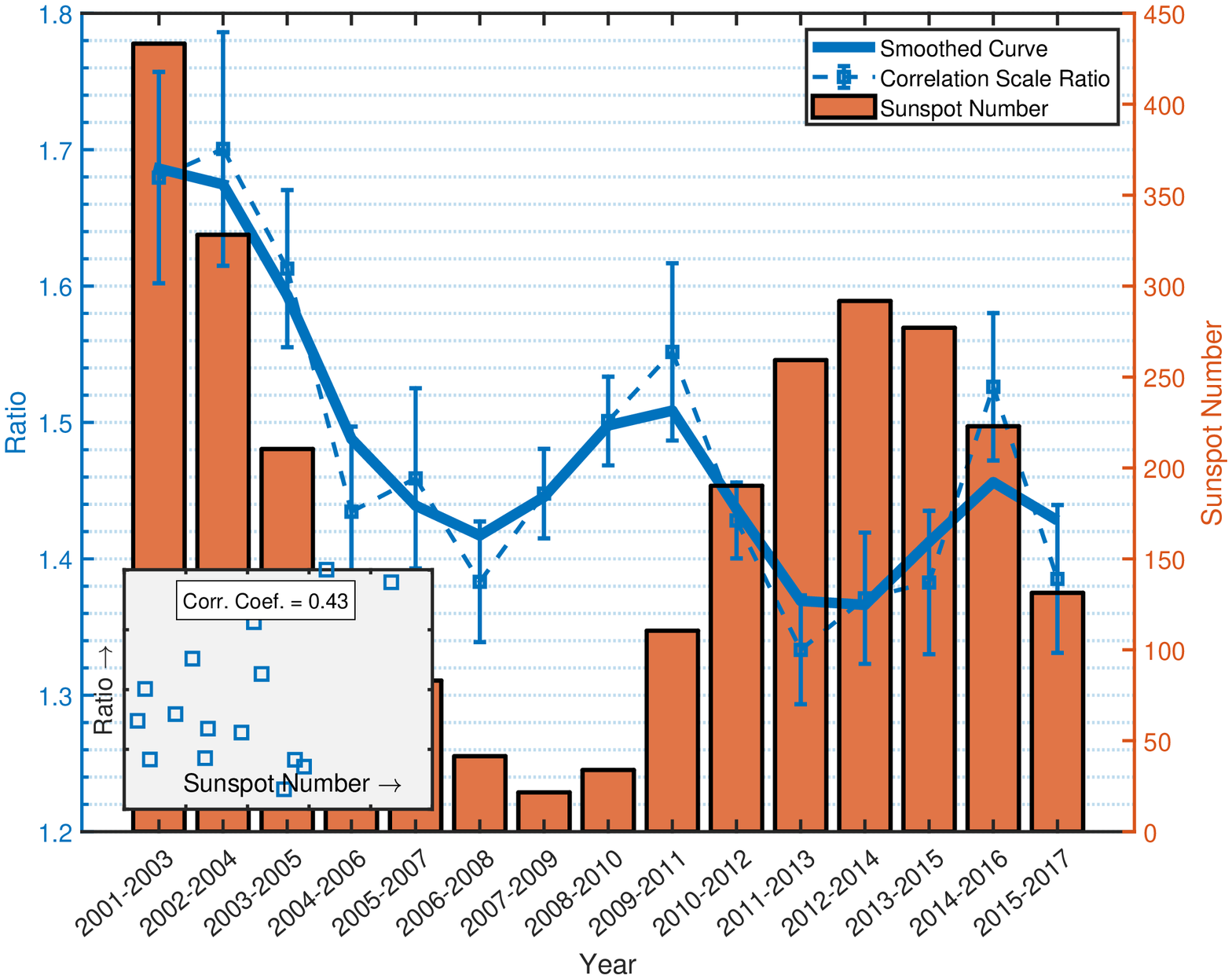}
 \caption{Top: distribution of the correlation length scales in different angular
 bins during the time period 2001-2017. The color scale bar on the right shows
 the values of the correlation scales. Bottom: evolution of the sunspot number
 and the ratios of the parallel to the perpendicular correlation scales during
 the period 2001-2017. The squares denote the ratios of the correlation
 scales, and the blue curve denotes the smoothed results of the ratios.
 The red histogram denotes the sunspot number in the divided time ranges.
 The inset shows the variations of the ratios of the correlation scales with
 the increase of the sunspot number. The correlation coefficient between the
 sunspot number and the anisotropy ratios of the correlation scales is 0.43. \label{3_cor_ratio}}
\end{figure}

The top panel of Figure \ref{3_cor_ratio} shows the distribution of
the correlation length scales in different angular bins during the
time period 2001-2017. The values of the correlation length scale
are determined from the robust fittings of the correlation function
with the exponential form as discussed in Section \ref{sec:method}.
As one can see, the correlation length scale decreases
systematically from the field-parallel to the field-perpendicular
directions. Based on these data, we further determine the ratio of
the parallel to the perpendicular correlation scales in each divided
time period, and calculate the smoothed results with the Gaussian
smoothing method for reducing the impact of extreme values. The
bottom panel of Figure \ref{3_cor_ratio} reveals the evolution of
the sunspot number and the ratio of the parallel to the
perpendicular correlation scales. The minimum ratio of the parallel
to the perpendicular correlation scales is 1.33, and the maximum
ratio is 1.70. The mean value of all the correlation scale ratios is
1.48, which is similar to the results provided in
\citet{2005ApJ...Dasso} and \citet{2007ApJ...654L.103O}. As we can
see, all the correlation scale ratios are larger than 1. This result
indicates that in the solar wind turbulence, the correlation length
scale is anisotropic, and the 2D component is dominant. The
correlation coefficient between the sunspot number and the
anisotropy ratio of the correlation scale is 0.43, which suggests
that there exists a moderately positive correlation between the
solar activity and the anisotropy of the correlation length scales.
Therefore, the influence of the solar activity on the anisotropy of
the correlation length scales is moderate.

\begin{figure}
 \epsscale{0.5}
 \plotone{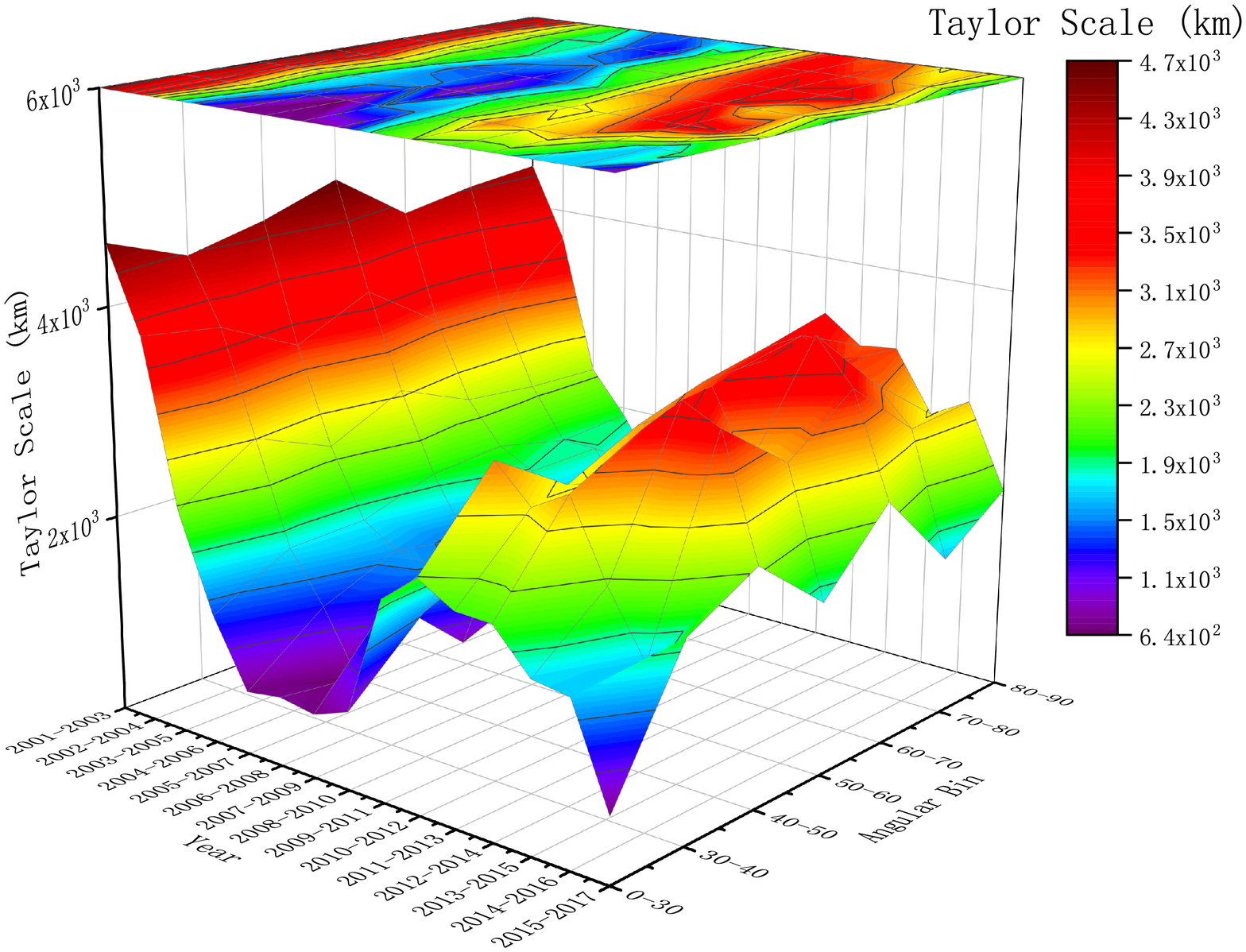}
 \plotone{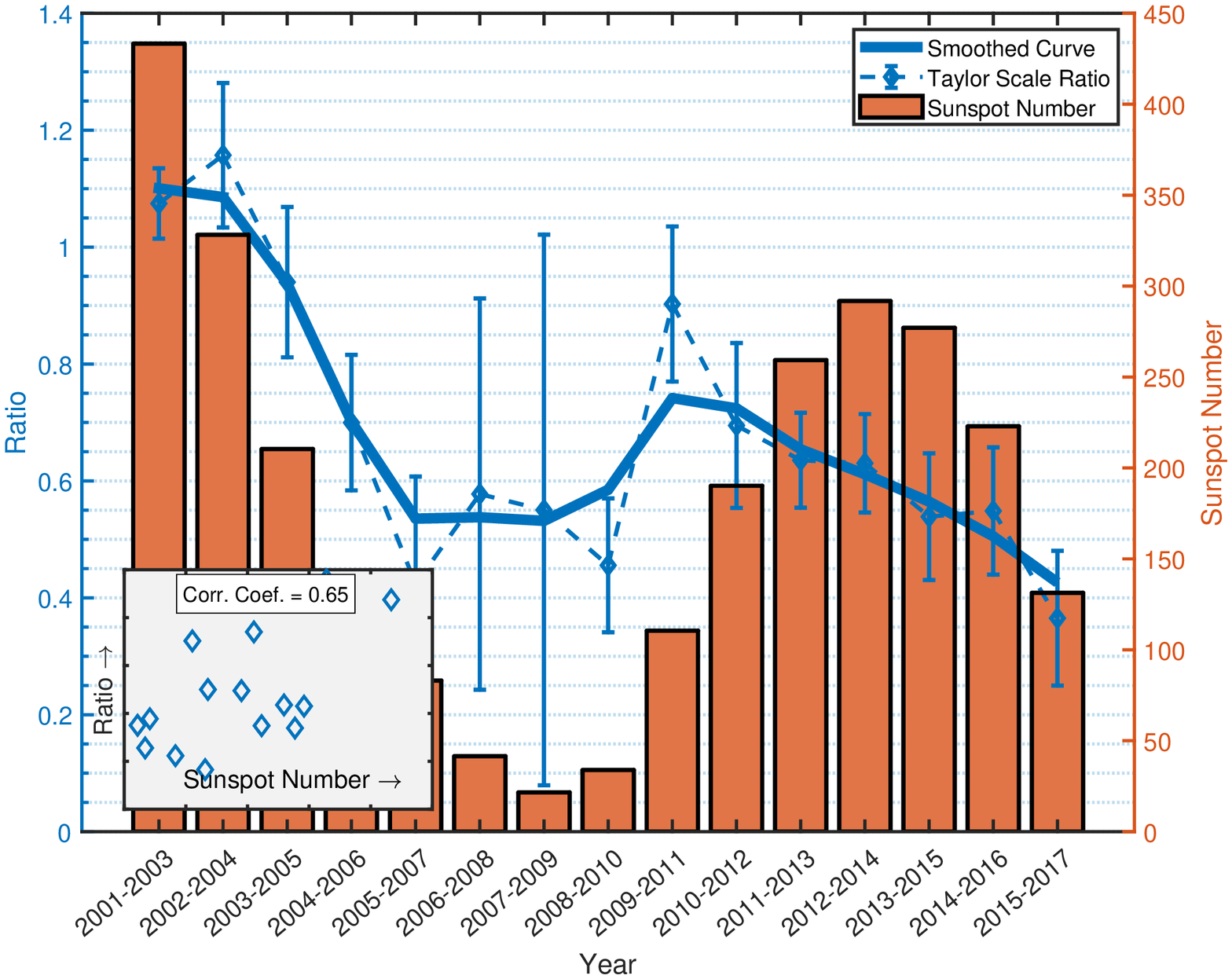}
 \caption{Top: distribution of the Taylor length scales in different angular
 bins during the time period 2001-2017. The color scale bar on the right
 shows the values of the Taylor scales. Bottom: evolution of the sunspot
 number and the ratios of the parallel to the perpendicular Taylor
 scales during the period 2001-2017. The diamonds denote the ratios
 of the Taylor scales, and the blue curve denotes the smoothed
 results of the ratios. The red histogram denotes the sunspot number
 in the divided time ranges. The inset shows the variations of the
 ratios of the Taylor scales with the increase of the sunspot
 number. The correlation coefficient between the sunspot number and
 the anisotropy ratios of the Taylor scales is 0.65. \label{3_tay_ratio}}
\end{figure}

The top panel of Figure \ref{3_tay_ratio} displays the distribution
of the Taylor length scales. The values of the Taylor length scale
in each angular bin and during every divided time period are
determined from the Richardson extrapolation method discussed in
\citet{2007JGRA..11210201W}. Dissimilar to the correlation length
scales shown in Figure \ref{3_cor_ratio}, the Taylor length scales
present more complex variations across the different angular bins.
During the years 2001-2004, the Taylor length scales along the mean
magnetic field are slightly larger than those along the
field-perpendicular directions. During the years 2004-2017, however,
the Taylor length scales are larger in the field-perpendicular
directions. In addition, the Taylor length scales in the
${40^\circ}-{70^\circ}$ angular bins seem to be relatively larger.
In general, the three-dimensional (3D) diagram of the Taylor length
scales displays a ``double-peak structure'' in some divided time
periods. Based on these data, we can determine the ratio of the
parallel to the perpendicular Taylor length scales, and can
calculate the smoothed results with the Gaussian smoothing method.
The bottom panel of Figure \ref{3_tay_ratio} presents the evolution
of the sunspot number and the ratio of the parallel to the
perpendicular Taylor length scales. The minimum of the ratios of the
parallel to the perpendicular Taylor scales is 0.37, and the maximum
is 1.16. The averaged value of all the Taylor scale ratios is 0.68,
which is similar to the result $0.91 \pm 0.45$ implied in
\citet{2009JGRA..114.7213W}. As shown, most of the Taylor scale
ratios are smaller than 1, which indicates that in the solar wind
turbulence, the Taylor length scale is generally anisotropic, and
the slab component is dominant. The correlation coefficient between
the sunspot number and the anisotropy ratio of the Taylor scale is
0.65. This result suggests that there exists a relatively strong
positive correlation between the solar activity and the anisotropy
of the Taylor length scales.

\begin{figure}
 \epsscale{0.5}
 \plotone{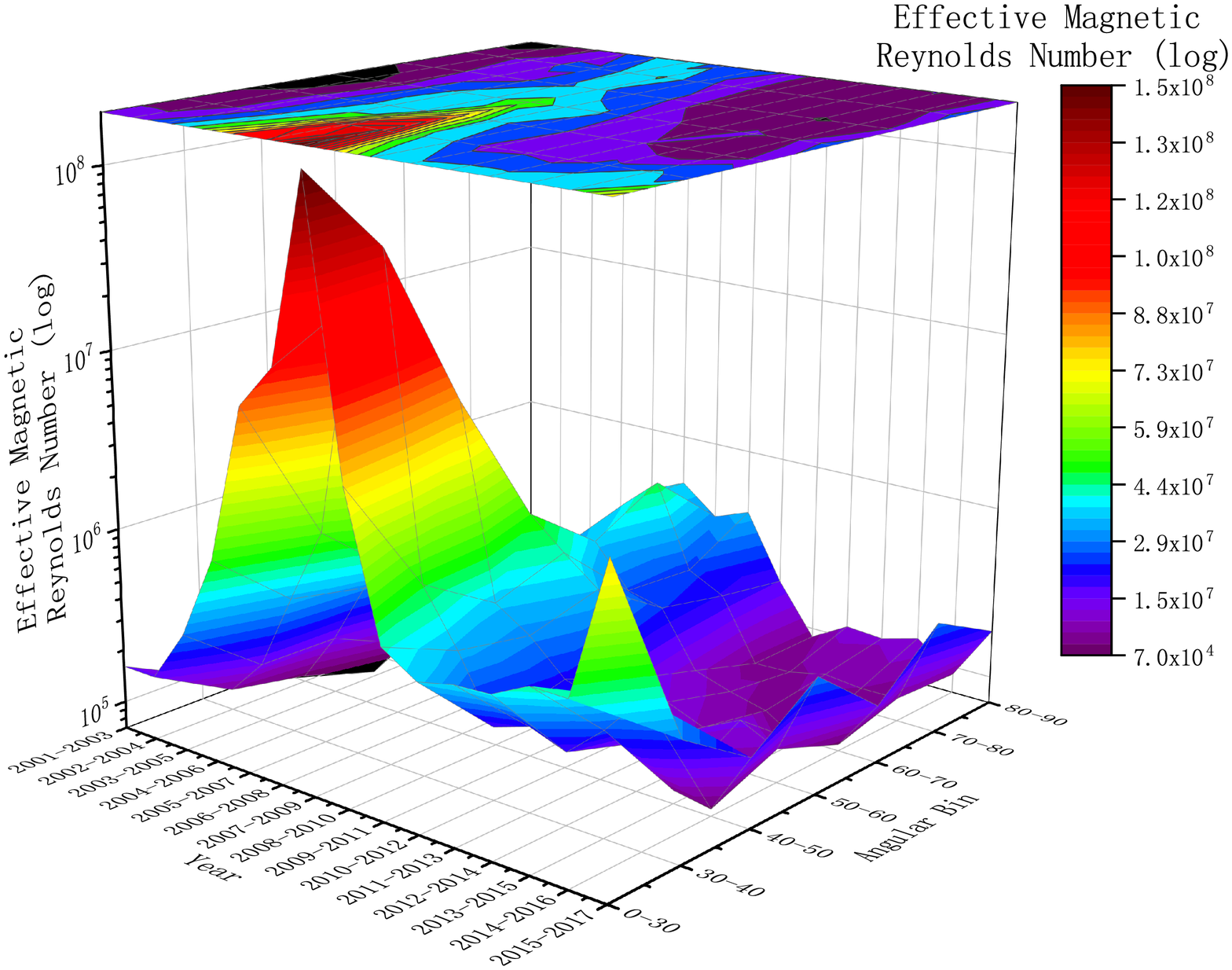}
 \plotone{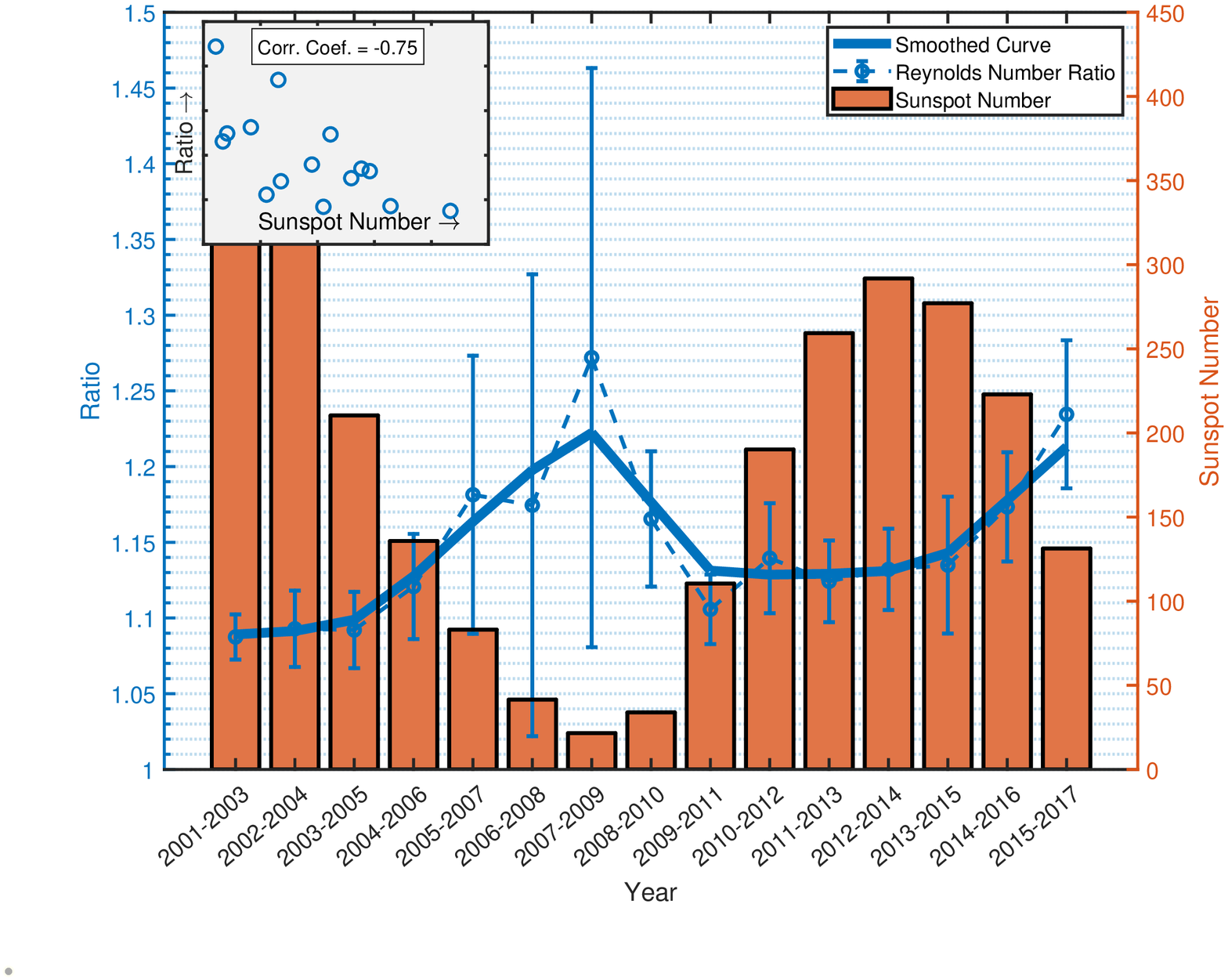}
 \caption{Top: distribution of the effective magnetic Reynolds
 number (in logarithmic unit) in different angular bins during the time
 period 2001-2017. The color scale bar on the right shows the values
 of the effective magnetic Reynolds number in logarithmic unit.
 Bottom: evolution of the sunspot number and the ratios of the parallel
 to the perpendicular effective magnetic Reynolds number (logarithmic
 unit) during the period 2001-2017. The circles denote the ratios of
 the effective magnetic Reynolds number, and the blue curve denotes
 smoothed results of the ratios. The red histogram denotes the sunspot
 number in the divided time ranges. The inset displays the variations of
 the ratios of the effective magnetic Reynolds number with the increase of the
 sunspot number. The correlation coefficient between the sunspot
 number and the anisotropy ratios of the effective magnetic Reynolds
 number is -0.75. \label{3_rey_ratio}}
\end{figure}

The top panel of Figure \ref{3_rey_ratio} presents the distribution
of the effective magnetic Reynolds number. The values of the
effective magnetic Reynolds number in each angular bin and during
each divided time period are calculated by using Equation \ref{R_m}.
As shown in the previous results of \citet{2020ApJ...899L..32Z}, the
effective magnetic Reynolds number is closely related with the solar
activity. During the period 2001-2017, the minimum and maximum
values of the effective magnetic Reynolds number are 96467.7 and
692071.1, respectively, and the averaged value of the effective
magnetic Reynolds number is 302827.4 \citep{2020ApJ...899L..32Z}.
For a clearer illustration, here we use a logarithmic unit in
presenting the results of the effective magnetic Reynolds number. As
one can see, the effective magnetic Reynolds number in the
field-parallel directions is systematically larger than that in the
field-perpendicular directions, which means that the 2D component is
approximately dominant. The bottom panel of Figure \ref{3_rey_ratio}
shows the evolution of the sunspot number and the ratio of the
parallel to the perpendicular effective magnetic Reynolds number
during the years 2001-2017. The minimum and maximum ratios of the
parallel to the perpendicular effective magnetic Reynolds numbers
are 1.09 and 1.27, respectively, and the mean value of all the
ratios of the effective magnetic Reynolds number is 1.15, which is
quite similar to the values of $\sim 1.14$ computed from the results
presented in \citet{2009JGRA..114.7213W}. The correlation
coefficient between the sunspot number and the anisotropy ratio of
the effective magnetic Reynolds number is -0.75, which indicates
that there exists a relatively strong negative correlation between
the solar activity and the anisotropy of the effective magnetic
Reynolds number. That is to say, the effective magnetic Reynolds
number is less anisotropic during a solar maximum.

\subsection{Discussion}
As shown in Figure \ref{3_cor_ratio}, the anisotropy of the
correlation length scale is only weakly affected by the solar
activity, namely, the anisotropy of the correlation scale is
relatively stable during the whole solar cycle. In addition, the
anisotropy of the correlation length scales is dominantly controlled
by the 2D component, which is consistent with the previous studies.
Due to the weak dependence on the solar activity, the anisotropy
values of the correlation scale investigated by different authors
during different phases of solar cycles usually reveal the similar
results.

Although the consistent results can be found for the anisotropy of
the correlation scale, this is not the case for the anisotropy of
the Taylor (and/or dissipation) scale. At the Taylor and dissipation
scales, the observational investigations of the correlation
anisotropy usually present inconsistent results. These conflicting
results in the literature can be reconciled by our findings in this
work. As shown in Figure \ref{3_tay_ratio}, the correlation
coefficient between the sunspot number and the ratio of the
field-parallel to the field-perpendicular Taylor length scales is
0.65, which indicates that the effects of the solar activity on the
correlation anisotropy near the Taylor scales are relatively strong.
Therefore, the anisotropy of the Taylor scale is relatively unstable
and varies with the declining and rise activity phases of solar
cycles. Specifically, the anisotropy of the Taylor scale is
2D-component dominated during the strong solar activity phases, and
is slab-component dominated during the weak solar activity phases.
This finding suggests that the long-term solar activity variations
may significantly affect the anisotropy of the Taylor scales. In
addition, the 3D diagram of the Taylor length scales in Figure
\ref{3_tay_ratio} displays a ``double-peak'' structure in some
divided time ranges. The Taylor length scales in the angular bins
${40^\circ}-{70^\circ}$ are relatively large. During the years
2011-2015 (solar maximum), this phenomenon is more pronounced.

The distribution of the correlation length scales does not show
obvious ``double-peak'' structure. Instead, it displays a relatively
small gradient in the angular bins ${40^\circ}-{70^\circ}$,
especially at the solar maximum. Combining the behaviors of the
correlation scales and the Taylor scales, we can find that both
these scales show some unusual properties near the angle of
${50^\circ}$. To specifically investigate this interesting
phenomenon, we calculate the correlation coefficients between the
sunspot number and both the correlation length scales and the Taylor
length scales in each angular bin, and present the results in Table
\ref{table}. As we can see, the values of the correlation
coefficients for the correlation scales decrease when the angular
bins vary from ${0^\circ}-{30^\circ}$ (approximately field-parallel)
to ${40^\circ}-{60^\circ}$. In the angular bins
${40^\circ}-{50^\circ}$ and ${50^\circ}-{60^\circ}$, the correlation
coefficients become very small (i.e., $0.04\pm0.09$ and
$0.05\pm0.05$). These nearly-zero values indicate that the
correlation scales in these directions are not affected by the solar
activity. However, the correlation coefficient in the angular bin
${0^\circ}-{30^\circ}$ (nearly field-parallel) is large (i.e.,
$0.57\pm0.03$). This result suggests that the correlation scale in
the field-parallel direction is considerably affected by the solar
activity. This finding will be useful for understanding the
anisotropy of the correlation scales. For the Taylor scales, the
values of the correlation coefficients generally increase when the
angular bins vary from ${0^\circ}-{30^\circ}$ (nearly
field-parallel) and ${80^\circ}-{90^\circ}$ (nearly
field-perpendicular) to ${50^\circ}-{60^\circ}$. Therefore, near the
angular bin ${50^\circ}-{60^\circ}$, where the value of the
correlation coefficient is $0.95\pm0.02$, the Taylor scale is most
easily influenced by the solar activity. We note that in all
directions relative to the mean magnetic field, the solar activity
significantly influences the Taylor scales. This is one
manifestation of the complexity of the Taylor scale anisotropy.

In this work, we do not investigate the specific effects of
fast/slow solar wind on the correlation scale anisotropy and the
Taylor scale anisotropy, which is also an interesting topic in the
field of solar wind turbulence and was investigated by
\citet{2011JGRA..116.8102W}. To compare with the results in
\citet{2011JGRA..116.8102W} and other relevant works in the context
of fast/slow solar wind may be the subject of our future work. In
this paper, we primarily investigate the solar cycle variations of
the anisotropy of Taylor scale and correlation scale. Note that we
mainly present the experimental observation results of the solar
cycle variations of the anisotropy in the solar wind turbulence. A
detailed discussion and explanation regarding the physical mechanism
of this phenomenon may be one topic of our future work.

\section{Summary} \label{sec:summary}
Based on the simultaneous two-point correlation function
measurements, in this work we investigate the solar cycle variations
of the anisotropy of the correlation scales and the Taylor scales in
the solar wind turbulence. The magnetic field data used in this
investigation were measured by the triaxial fluxgate magnetometers
on board spacecraft ACE, Wind, and Cluster during the period from
January 2001 to December 2017, which covers more than an entire
solar cycle. The data accumulated over a long time are sufficient to
study the effects of long-term solar activity on the anisotropy of
the solar wind turbulence. Generally, the correlation scale length
decreases with the relative angle between magnetic field fluctuation
and the average magnetic field, independent of the solar activity.
The averaged value of all the ratios of the correlation scales is
1.48, which is similar to the results presented in
\citet{2005ApJ...Dasso} and \citet{2007ApJ...654L.103O}. The
correlation coefficient between the sunspot number and the
anisotropy ratios of the correlation scales is 0.43, which indicates
that the influence of the solar activity on the anisotropy of the
correlation length scales is not so significant. Furthermore, we
find that the anisotropy of the correlation scales is dominated by
the 2D component. However, this is not the case for the anisotropy
of the Taylor (and/or dissipation) scales. The minimum ratio of the
parallel to the perpendicular Taylor scales is 0.37, and the maximum
ratio is 1.16. The averaged value of all the ratios is 0.68, which
is consistent with the values of $0.91 \pm 0.45$ calculated from the
results in \citet{2009JGRA..114.7213W}. The correlation coefficient
between the sunspot number and the anisotropy ratios of the Taylor
scales is 0.65, which indicates that the anisotropy of the Taylor
scales is relatively significantly affected by the long-term solar
activity. We further find that near/during the solar maximum, the
anisotropy of the Taylor scale is 2D-component dominated, while
near/during the solar minimum, the anisotropy of the Taylor scale is
slab-component dominated.

Using the Taylor scales and the correlation scales, we further
determine the value of the effective magnetic Reynolds number for
each angular bin. We find that the effective magnetic Reynolds
number in the field-parallel directions is systematically larger
than that in the field-perpendicular directions, which indicates
that the 2D component is dominant. The averaged value of all the
ratios of the effective magnetic Reynolds number is 1.15, which
agrees well with the values of $\sim 1.14$ calculated from the
results shown in \citet{2009JGRA..114.7213W}. The correlation
coefficient between the sunspot number and the anisotropy ratios is
-0.75, which implies that there exists a relatively strong negative
correlation between the solar activity and the anisotropy of the
effective magnetic Reynolds number.

In addition, the correlation coefficient between the sunspot number
and the Taylor length scales is systematically larger than that
between the sunspot number and the correlation length scales. This
result means that the Taylor scale is more easily influenced by the
long-term solar activity than the correlation scale. Furthermore,
the correlation scales near the field-parallel and
field-perpendicular directions are most easily affected by the solar
activity among all the directions. However, the Taylor scales in all
directions relative to the mean magnetic field are significantly
influenced by the long-term solar activity. These results will be
very useful for better understanding the anisotropy of the solar
wind turbulence and especially its solar-cycle variability.


\acknowledgments

This work was supported in part by the B-type Strategic Priority
Program of the Chinese Academy of Sciences under grant XDB41000000,
the National Natural Science Foundation of China under grants
41874207, 41621063, 41474154, and 41204130, and the Chinese Academy
of Sciences under grant KZZD-EW-01-2. H.-Q.H. gratefully
acknowledges the partial support of the Youth Innovation Promotion
Association of the Chinese Academy of Sciences (No. 2017091). We
benefited from the data of ACE, Wind, and Cluster provided by
NASA/Space Physics Data Facility (SPDF)/CDAWeb. The sunspot data
were provided by the World Data Center SILSO, Royal Observatory of
Belgium, Brussels.


\clearpage


\begin{sidewaystable}[htbp]
    \centering
    \caption{Values of correlation coefficients between sunspot number and correlation length scales and Taylor length scales in different angular bins during 2001-2017. \label{table}}
    \newsavebox{\tablebox}
    \begin{lrbox}{\tablebox}
    \begin{tabular}{c|c|c|c|c|c|c|c}
        \hline
        &${0^\circ-30^\circ}$&${30^\circ-40^\circ}$&${40^\circ-50^\circ}$&${50^\circ-60^\circ}$&${60^\circ-70^\circ}$&${70^\circ-80^\circ}$&${80^\circ-90^\circ}$ \\
        \hline
        Correlation scale&$0.57\pm0.03$&$0.31\pm0.11$&$0.04\pm0.09$&$0.05\pm0.05$&$0.56\pm0.05$&$0.65\pm0.05$&$0.34\pm0.09$ \\
        \hline
        Taylor scale&$0.84\pm0.03$&$0.81\pm0.05$&$0.91\pm0.03$&$0.95\pm0.02$&$0.88\pm0.05$&$0.89\pm0.03$&$0.90\pm0.02$ \\
        \hline
    \end{tabular}
    \end{lrbox}
    \resizebox{1.0\textwidth}{!}{\usebox{\tablebox}}
\end{sidewaystable}


\clearpage

\clearpage

\clearpage

\clearpage


\end{document}